\documentclass{emulateapj}
\usepackage{apjfonts}
\lefthead{HAN} \righthead{LENS-LIGHT ANALYSIS METHOD}

\newcommand{\vvec}{\mbox{\boldmath $v$}}

\newcommand{\te}{t_{\rm E}}
\newcommand{\re}{r_{\rm E}}

\newcommand{\retilde}{\tilde{r}_{\rm E}}
\newcommand{\thetae}{\theta_{\rm E}}

\begin{document}
\title{On the Feasibility of Characterizing Lens Stars in 
Future Space-Based Microlensing Surveys}


\author{Cheongho Han}
\affil{Department of Physics, Institute for Basic Science
Research, Chungbuk National University, Chongju 361-763, Korea;
cheongho@astroph.chungbuk.ac.kr}


\begin{abstract}

If a light-emitting star is responsible for a gravitational microlensing 
event, the lens can be characterized by analyzing the blended light from 
the lens.  In this paper, we investigate the feasibility of characterizing 
lenses by using this method in future space-based lensing surveys.  To 
judge the feasibility of the method, we estimate the portions of events 
whose blended flux $F_b$ can be firmly noticed and most of it can be 
attributed to the lens by carrying out detailed simulations of Galactic 
bulge lensing events considering various blending sources, including 
the lens, background stars, and binary companions to the lens and source.  
From this, it is estimated that among the events to be detected from a 
survey using a 1 m space telescope, $\sim 27\%$ will have blending 
fractions of $F_b/F\geq 10\%$ and the blended flux of half of these 
events will be contaminated more than $(F_b-F_L)/F_b=20\%$ by the flux 
from blending sources other than the lens, implying the contamination 
of the blended flux will be substantial.  Although the contamination by 
the background stars can be reduced by using an instrument with a higher 
resolution, it is estimated that the blended flux of more than 1/3 of 
events will still be contaminated (mostly by binary companions) even 
using a telescope equivalent to the {\it Hubble Space Telescope}, assuming 
50\% binary frequency.  We, therefore, conclude that caution and 
consideration of the blending contaminants are required in applying the 
lens-light analysis method.

\end{abstract}


\section{Introduction}

Proposed by \citet{paczynski86}, initiated by the MACHO 
\citep{alcock93, alcock95}, EROS \citep{aubourg93}, and OGLE 
\citep{udalski94} collaborations, and succeeded by the OGLE-II 
\citep{wozniak01}, OGLE-III \citep{udalski03}, and MOA \citep{bond01} 
collaborations, the Galactic microlensing experiments have been and 
are searching for microlensing events toward the Galactic bulge and 
Magellanic Cloud fields.  Current experiments are routinely detecting 
more than 500 events every season and the total number of detected 
events now exceeds 2000.  Except for several dozens of events, most of 
these events have been detected toward the Galactic bulge field.

The light curve of a microlensing event is represented by
\begin{equation}
F = AF_0 + F_b;\qquad A={u^2+2\over u(u^2+4)},
\label{eq1}
\end{equation}
where $F_0$ is the baseline flux of the source star, $F_b$ is the 
blended flux, $A$ represents the lensing magnification, and $u$ is 
the lens-source separation normalized by the angular Einstein radius 
$\thetae$.  When the  lensing event is observed, one can measure 
three observables related to the physical parameters of the lens, 
namely the Einstein timescale $\te$, the angular Einstein radius 
$\thetae$, and the Einstein ring radius projected onto the observer 
plane $\retilde$.  These observables are related to the underlying 
physical lens parameters of the mass $M$, relative lens-source 
parallax $\pi_{\rm rel}={\rm AU}\ (D_{\rm L}^{-1}- D_{\rm S}^{-1})$, 
and proper motion $\mu_{\rm rel}$, by \citep{gould00b}
\begin{equation}
\te = {\thetae\over \mu_{\rm rel}},\ \ \ 
\thetae=\sqrt{4GM\pi_{\rm rel} \over c^2\ {\rm AU}},\ \ \ 
\retilde = \sqrt{4GM\ {\rm AU} \over c^2 \pi_{\rm rel}}.
\label{eq2}
\end{equation}
Once all these observables are measured, the lens mass is 
uniquely determined by
\begin{equation}
M=\left( {c^2 \over 4G} \right)\retilde \thetae.
\label{eq3}
\end{equation}
Among these three observables, however, only $\te$ is routinely 
measurable from the lensing light curve and $\thetae$ and $\retilde$ 
can be measured only in some favorable cases.  As a result, of the 
roughly 2000 microlensing events detected to date, there have been 
only about a dozen events for which $\thetae$ has been measured and 
another dozen for which $\retilde$ has been measured.  Moreover, 
simultaneous measurements of $\thetae$ and $\retilde$ are very 
rare, and so far there are only three events, MACHO-LMC-5, 
OGLE-2000-BLG-5, and OGLE-2003-BLG-238, for which the microlens
 mass and distance have been reliably determined \citep{alcock01, 
an02, jiang04, gould04}.

If a light-emitting star is responsible for a microlensing event, the 
lens can be characterized by another method.  In this case, the light 
from the lens star contributes to the observed flux and thus one can 
characterize the lens star by analyzing the flux from the lens 
\citep{kamionkowski95, buchalter96, mao98}: `lens-light analysis' 
method.  If realized, this method can be applied to a large number of 
events because it is believed that a majority of events detected toward 
the Galactic bulge field are caused by stars \citep{han03}.  Currently, 
this method is not being used due to two major observational difficulties.  
First, because of the high number density of source stars in the Galactic 
bulge field combined with the poor resolution of the ground-based observation, 
current lensing surveys can effectively monitor only bright resolved stars,
and thus the relative contribution of the lens flux to the total 
observed flux is small.  Second, even if the source star is relatively 
faint, most blended flux comes from nearby background stars, and thus 
the flux from the lens has a meager contribution to the total blended 
flux.

However, the situation might be different in future space-based 
microlensing surveys, such as the {\it Galactic Exoplanet Survey Telescope} 
({\it GEST}) mission \citep{bennett02}, whose concept was later 
succeeded by {\it Microlensing Planet Finder} ({\it MPF}) mission
\citep{bennett04}.  The proposed mission plans to monitor faint 
main-sequence stars to optimize the detections of terrestrial planetary 
signals by minimizing finite-source effect.  In addition, events to be 
observed by the mission would suffer from much less blending caused by 
background stars thanks to the high resolution provided by space 
observation.  Then, the proposed space-based microlensing survey might 
be able to detect enough light from the lens stars for a significant 
fraction of the event sample.  By measuring the apparent magnitude and 
color of the lens star, then, it would be possible to determine the 
spectral type of the lens star and thus can estimate the approximate 
lens mass and distance.

The feasibility of the lens-light analysis method depends greatly on 
whether the flux from the lens accounts for not only a substantial
fraction of the observed flux but also most of the blended flux.  If the 
blended flux is seriously contaminated by the flux from blending sources 
other than the lens, analysis of the blended light would be difficult or 
might lead to wrong characterization of the lens star.  Despite the high 
resolution of the space mission, accidental overlap of the source star 
image with those of background stars will be unavoidable for a fraction 
of events.  In addition, binary companions to either the source star or 
the lens can also contaminate the blended flux.  Therefore, it is 
important to evaluate the degree of contamination to the blended flux 
by these possible contaminants.  In this paper, we judge the feasibility 
of the lens-light analysis method in future space-based microlensing 
surveys by estimating the fractions of events where the blended flux 
can be firmly noticed and the most of the blended flux is attributed 
to the lens star.

\section{Microlensing Event Simulations}

To estimate the fraction of events whose blended flux can be noticed 
and attributed mostly to the lens star, we carry out detailed simulations 
of Galactic bulge microlensing events considering blending caused by 
various sources, including the lens star, background stars (denoted as 
`bs'), and the companions to the source and lens (denoted as `S2' and 
`L2', respectively).

The simulations of lensing events are proceeded as follows.
\begin{enumerate}
\item
We assume that the survey is conducted toward the Baade's Window with 
the central coordinates of $(l,b)=(1^\circ,-4^\circ)$.  Based on the
luminosity function (LF) of \citet{holtzman98} constructed using the 
{\it Hubble Space Telescope} ({\it HST}),  we normalize the stellar number 
density to $1.2\times 10^4/{\rm arcmin}^2$ for stars up to $M_V=12.25$, 
which corresponds to the brightness of a M5 star.  Blending by background 
stars is taken into consideration by assuming that the images of stars 
cannot be resolved if the separation between stars is less than the 
the angular resolution, $\theta_{\rm res}$.  To see the dependence on 
the angular resolution, we test two cases of $\theta_{\rm res}=
0''\hskip-2pt .1$ and $0''\hskip-2pt .25$, which correspond to the 
resolutions of the {\it HST} (with an aperture of ${\cal D}=2.4\ {\rm m}$) 
and {\it MPF} (with  ${\cal D}=1.0\ {\rm m}$), respectively.

\item
For the density and velocity distributions of the lens and source, 
we adopt the barred bulge model of \citet{han03}.  In the model, the 
bulge mass distribution is scaled by the deprojected infrared light 
density profile of \citet{dwek95}, specifically model G2 with 
$R_{\rm max}=5$ kpc in their Table 1.  The velocity distribution of 
the bulge is deduced from the tensor virial theorem and the resulting 
distribution of the lens-source transverse velocity, $\vvec$, is 
listed in Table 1 of \citet{han03}, specifically non-rotating barred 
bulge model.  In our simulations, we consider only bulge self-lensing 
events because they account for the majority of Galactic bulge events
\citep{zhao96}.

\item
The absolute magnitude of the source star is assigned based on the 
binary-corrected $V$-band LF of stars in the solar neighborhood listed 
in \citet{allen00}.\footnote{For the source brightness, it might be 
thought that using the LF of Galactic bulge stars obtained from the 
{\it HST} observation would be a better choice.  However, this LF is 
not binary-corrected because the companions to source stars cannot be 
resolved even with the {\it HST}.  Since the purpose of this paper is 
investigating the effect of contaminants to the blended flux and the 
binary companion to the source star is an important contaminant candidate, 
we use binary-corrected LF of solar-neighborhood stars instead of that 
of binary-corrected LF of Galactic bulge stars.} We assume that the 
brightness range of the source star to be monitored by the survey is 
$2.7\lesssim M_V\lesssim 7.5$, which corresponds to early F to late 
K-type main-sequence stars.  The apparent source magnitude is determined 
considering the distance to the source star and extinction.  The extinction 
is determined such that the source star flux decreases exponentially with 
the increase of the dust column density, where the dust column density is 
computed based on an exponential dust distribution model with a scale 
height of $h_z=120\ {\rm pc}$, i.e.\ $\propto \exp(-|z|/h_z)$.  We 
normalize the amount of extinction such a way that  that $A_V=1.28$ 
for a star located at $D_{\rm S}=8\ {\rm kpc}$ following the measurement 
of \citet{holtzman98}.  The brightness of the companion to the source 
star is assigned under the assumption that the companion follows the 
same LF of the primary star \citep{duquennoy91}.  Since the star to be 
monitored by the survey will be selected based on the combined flux of 
the primary and companion, there is no lower limit of the companion 
star brightness.  We note, however, that the terms `primary' does not 
designate the brighter component of the binary, but it indicates the 
star participating in the lensing magnification, although the primary 
source star is brighter than the companion in most cases.

\item
We assign the lens mass based on the mass function (MF) of 
\citet{gould00a}.  The model MF is composed of stars following a double 
power-law distribution, brown dwarfs (BDs), and stellar remnants of 
white dwarfs (WDs), neutron stars (NSs), and black holes (BHs), where 
the mass fractions of the individual lens components are
\begin{equation}
{\rm stars}:{\rm BD}:{\rm WD}:{\rm NS}:{\rm BH}=62:7:22:6:3.
\label{eq4}
\end{equation}
Once the mass of the lens is chosen and it is turned out to be a 
stellar lens, its brightness is assigned based on the mass by using 
the mass-$M_V$ relation listed in \citet{allen00}.  If the lens is 
a remnant, on the other hand, it is assumed to be dark.  The brightness 
of the companion to the lens is assigned by the same way as the primary 
lens.  Similar to the notations for the source stars, the primary lens 
designates the binary component that is involved with the lensing process.  
The apparent magnitude of the lens is determined by the same way as that 
of the source star considering the distance and extinction.

\item
Once the lens and source are chosen, the rate of the event associated 
with the lens and source is computed by
\begin{equation}
\Gamma \propto \rho(D_{\rm S})D_{\rm S}^2 \rho(D_{\rm L}) \sigma v 
u_{0,{\rm th}},
\label{eq5}
\end{equation}
where $\rho(D)$ is the matter density along the line of sight, the 
factor $D_{\rm S}^2$ is included to account for the increase of the 
number of source stars with the increase of $D_{\rm S}$, $\sigma$ 
represents the lensing cross-section corresponding to the diameter 
of the Einstein ring, i.e.\ $\sigma=2\re=2D_{\rm L}\thetae$, and 
$u_{0,{\rm th}}$ represents the threshold lens-source separation 
(normalized by $\thetae$) required for the event detection.  Due 
to blending, the threshold magnification for the event detection 
becomes higher.  The increased threshold magnification is 
\begin{equation}
{A'}_{\rm th}=A_{\rm th}\left( 1+{F_b\over F_{\rm S}} \right)-
{F_b\over F_{\rm S}},
\label{eq6}
\end{equation}
where $A_{\rm th}$ is the threshold magnification without blending.  
Then, the threshold lens-source separation corresponding to the 
increased threshold magnification \citep{distefano95, han99} is 
\begin{equation}
{u}_{0,{\rm th}} = \left[ {2\over \left( 1-{A'}_{\rm th}^{-2} 
\right)^{1/2}}  -2 \right]^{1/2}.
\label{eq7}
\end{equation}
In our simulations, we set $A'_{\rm th}=3/\sqrt{5}$ following the 
conventional threshold magnification.  Since the future lensing 
survey will be carried out with a high monitoring frequency, we 
assume that there is no bias in event detections depending on 
the event timescale.

\end{enumerate}

\section{Blending and Contamination to the Blended Flux}

We produce a large number ($10^5$) of events from the simulations 
following the procedure described in the previous section and 
compute the relative event rates of the individual events.  Based 
on these events and their rates, we then construct the distribution 
of the blended light fraction $F_b/F$.  In order to inspect the 
contamination of the blended flux by the blending sources other 
than the lens star, we also construct the distribution of the 
fraction of the lens flux $F_{\rm L}$ out of the total blended 
flux, $F_{\rm L}/F_b$, among the events for which blending is 
firmly detected.  We assume that blended light can be noticed 
if $F_b/F\geq 0.1$.  As candidates of the blending contaminants, 
we consider background stars, and companions to the source and lens, 
i.e.\ $F_b=F_{\rm L}+F_{\rm bs}+F_{\rm S2}+ F_{\rm L2}$, where 
$F_{\rm bs}$, $F_{\rm S2}$, and $F_{\rm L2}$ represent the fluxes 
from the background stars and companions to the source and lens, 
respectively.

\begin{figure}[th]
\epsscale{1.2}
\plotone{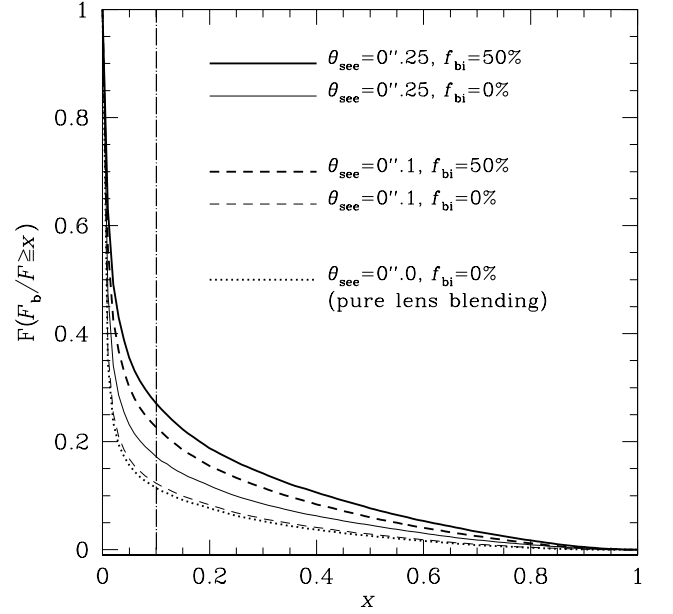}
\caption{\label{fig:one}
The portion of the Galactic bulge events where the blended light 
fraction, $F_b/F$, is greater than a certain value $x$. 
}\end{figure}

\begin{figure}[th]
\epsscale{1.2}
\plotone{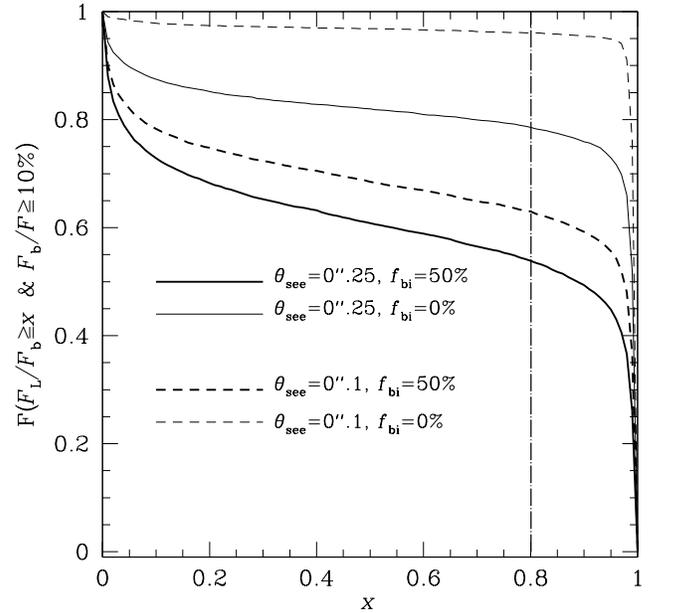}
\caption{\label{fig:two}
The portion 
of the Galactic bulge events where the lens light fraction out of the 
total blended flux, $F_{\rm L}/F_b$, is greater than a certain value 
$x$ among the blended events with $F_b/F\geq 10\%$.
}\end{figure}

In Figure~\ref{fig:one}, we present the distribution of the portion of 
events where the blended light fraction is greater than a certain value 
$x$, ${\cal F}( F_b/F\geq x)$.  In Figure~\ref{fig:two}, we also present 
the distribution of the portion of events where the lens light fraction 
out of the total blended flux is greater than a certain value $x$ among 
the blended events with $F_b/F\geq 0.1$, ${\cal F}( F_{\rm L} /F_b\geq 
x\ {\rm and}\ F_b/F\geq 0.1)$.  We test two binary frequencies of 
$f_{\rm bi}=0.5$ and $0.0$, where the latter corresponds to the case 
that the background star is the only blending contaminant, i.e.\ $F_b=
F_{\rm L}+F_{\rm bs}$.  To see the blending contribution only by the 
lens star, we also construct the distribution ${\cal F}(F_b/F\geq x)$ 
under the condition that $\theta_{\rm res} =0''\hskip-2pt.01$ and 
$f_{\rm bi}=0\%$, i.e.\ $F_b=F_{\rm L}$ (pure lens blending case).  
In Table~\ref{table1}, we summarize the results of blending and 
contamination of the blended flux.

\begin{deluxetable*}{crcccl}
\tablecaption{Blended Event Portions \label{table1}}
\tablewidth{0pt}
\tablehead{
\multicolumn{2}{c}{cases} &
\multicolumn{2}{c}{portions} &
\multicolumn{2}{c}{comments} \\
\colhead{$\theta_{\rm res}$} &
\colhead{$f_{\rm bi}$} &
\colhead{${\cal F}(F_b/F\geq 0.1)$} &
\colhead{${\cal F}(F_{\rm L}/F_b\geq 0.8\ \&\ F_b/F\geq 0.1)$} &
\colhead{instrument} &
\colhead{blending sources}}
\startdata
$0''\hskip-2pt .25$  & 50\%  &  27\%   &  53\%  & {\it MPF}  & L, bs, S2, L2 \\
      --             &  0\%  &  17\%   &  78\%  & --          & L, bs           \\
$0''\hskip-2pt .10$  & 50\%  &  23\%   &  63\%  & {\it HST}   & L, bs, S2, L2   \\
      --             &  0\%  &  13\%   &  97\%  & --          & L, bs           \\
$0''\hskip-2pt .01$  &  0\%  &  12\%   & 100\%  &             & L               \\
\enddata
\tablecomments{ 
The portion of events where the blended light fraction is greater 
than 10\%, ${\cal F}(F_b/F\geq 0.1)$, and the portion of events 
where the lens light fraction out of the total blended flux is 
greater than 80\% among the blended events with $F_b/F\geq 0.1$, 
${\cal F} (F_{\rm L}/F_b\geq 0.8\ \&\ F_b/F\geq 0.1)$. The values 
are estimated  under various angular resolutions $\theta_{\rm res}$ 
of the instrument and assumptions of the binary frequency $f_{\rm bi}$.
}
\end{deluxetable*}

From Table~\ref{table1} and Figure~\ref{fig:one} and \ref{fig:two}, 
we find the following results.
\begin{enumerate}
\item
From the distribution of  ${\cal F}( F_b/F\geq x)$ for the 
pure-lensing blending case ($\theta_{\rm res}=0''\hskip-2pt.01$ 
and $f_{\rm bi}=0\%$), one finds that the lens does contribute 
to the observed flux  for a substantial portion of events.  We 
find that the flux from the lens star will contribute to the 
observed flux more than $F_b/F\geq 0.1$ for $\sim 12\%$ of events.

\item
However, we find that the contribution to the blended flux by sources 
other than the lens star is not negligible as well.  Even not considering 
binary companions ($f_{\rm bi}=0\%$), we estimate that the blended flux 
will be contaminated by the flux from background stars by more than 
$(F_b-F_{\rm L})/F_b=20\%$ for nearly a quarter of events with 
$F_b/F\geq 0.1$ to be detected from the survey using a 1 m space 
telescope.
 
\item
Although the contamination by the background stars can be reduced by 
using an instrument with a higher resolution, the contamination by the 
binary companions to the source and lens is unavoidable.  When these 
additional contaminants are considered, we estimate that the portion 
of events that will suffer from relatively less severe blending 
($F_{\rm L}/ F_{b}\geq 0.8$) will be only 53\% and 63\% for surveys 
using instruments equivalent to the {\it MPF} and {\it HST}, respectively. 

\end{enumerate}

\section{Conclusion}

We investigated the feasibility of characterizing lens stars from the 
analysis of the blended lens flux in the future space-based lensing 
surveys.  We judged this feasibility by estimating the portions of 
events whose blended flux can be firmly noticed and most of the blended 
flux can be attributed to the lens star based on detailed simulations 
of bulge lensing events considering various blending sources, including 
the lens, background stars, and the binary companions to the lens and 
source.  From this estimation, we found that although the flux from 
the lens star will contribute to the observed flux for a substantial 
fraction of events, the contamination of the blended flux by the light 
from the blending sources other than the lens star will not be negligible.  
Considering these blending contaminants, we estimated that that among 
the events to be detected from a survey using a 1 m space telescope, 
$\sim 27\%$ will have blending fractions of $F_b/F\geq 10\%$ and the 
blended flux of half of these events will be contaminated more than 
$(F_b-F_L)/F_b=20\%$ by the flux from blending sources other than the 
lens.  Although the contamination by the background stars can be reduced 
by using an instrument with a higher resolution, it is estimated that 
the blended flux of more than 1/3 of events will still be contaminated 
(mostly by binary companions) even using a telescope equivalent to 
{\it HST}, assuming 50\% binary frequency.  We, therefore, conclude 
that caution and consideration of the blending contaminants are required 
in applying the lens-light analysis method

\acknowledgments 
This work was supported by the Astrophysical Research Center for the 
Structure and Evolution of the Cosmos (ARCSEC") of Korea Science and 
Engineering Foundation (KOSEF) through Science Research Program 
(SRC) program.

\end{document}